\documentclass[amsmath,floatfix,twocolumn,superscriptaddress]{revtex4-1}
\usepackage{subfigure}
\usepackage{siunitx}
\usepackage{amssymb}
\usepackage{amsmath}
\usepackage{graphicx}
\usepackage{array}
\usepackage{dcolumn}
\usepackage{psfrag}
\usepackage{bm}
\usepackage{color}
\usepackage{multirow}

\begin{document}

\title{Machine Learning Modeling of Temperature-Dependent Optoelectronic Properties of Anharmonic Solid Solutions}

\author{Pol Benítez}
\email{pol.benitez@upc.edu}
    \affiliation{Department of Physics, Universitat Politècnica de Catalunya, 08019 Barcelona, Spain}
    \affiliation{Research Center in Multiscale Science and Engineering, Universitat Politècnica de Catalunya,
    08019 Barcelona, Spain}

\author{Cibrán López}
    \affiliation{Department of Physics, Universitat Politècnica de Catalunya, 08019 Barcelona, Spain}
    \affiliation{Research Center in Multiscale Science and Engineering, Universitat Politècnica de Catalunya,
    08019 Barcelona, Spain}

\author{Edgardo Saucedo}
    \affiliation{Research Center in Multiscale Science and Engineering, Universitat Politècnica de Catalunya,
    08019 Barcelona, Spain}
    \affiliation{Departament d'Enginyeria Electrònica, Universitat Politècnica de Catalunya, 08034 Barcelona, Spain}

\author{Claudio Cazorla}
\email{claudio.cazorla@upc.edu}
    \affiliation{Department of Physics, Universitat Politècnica de Catalunya, 08019 Barcelona, Spain}
    \affiliation{Research Center in Multiscale Science and Engineering, Universitat Politècnica de Catalunya,
    08019 Barcelona, Spain}
    \affiliation{Institució Catalana de Recerca i Estudis Avançats (ICREA), Passeig Lluís Companys 23, 08010 Barcelona, Spain}

\begin{abstract}
\textbf{Abstract.}~Leveraging strong optoelectronic responses to external stimuli, such as temperature and electric fields, is 
central to the development of advanced photonic technologies, including adaptive photodetectors and reconfigurable photovoltaic 
devices. However, only a limited number of semiconducting materials, typically characterized by strong electron–phonon coupling, 
are known to exhibit such pronounced responsiveness, and their equilibrium optoelectronic properties are often not optimally 
suited for targeted applications. Chemical engineering strategies, such as doping and solid-solution mixing, are therefore widely 
employed to fine-tune the electronic and optical properties of semiconductors. Predicting the impact of such modifications, however, 
remains highly challenging due to the intrinsic complexity of chemically disordered and anharmonic systems, as well as the 
computational limitations of conventional first-principles approaches. In this work, we introduce a new computational framework 
that combines \textit{ab initio} electronic-structure methods with machine-learning techniques to achieve first-principles 
precision in the prediction of optoelectronic properties of anharmonic solid solutions at finite temperature. We apply this 
approach to Ag$_{3}$SBr$_{x}$I$_{1-x}$ solid solutions, an emergent class of optoelectronic materials that have been experimentally 
shown to exhibit large band-gap tunability and strong responses to thermal excitations. Our results provide quantitative insight 
into the interplay between chemical disorder, lattice dynamics, and electronic structure in these materials. More broadly, this 
study establishes a general strategy for the accurate modeling of optoelectronic functionality in chemically disordered semiconductors.
\\

\end{abstract}

\maketitle

\section{Introduction}
\label{sec:intro}
Semiconducting materials play a central role in contemporary and emerging energy and optoelectronic technologies, including photovoltaics 
and integrated electronic circuits. Their functional behaviour is dictated by the electronic band structure, which in many systems is not 
a static quantity but evolves with temperature due to electron–phonon coupling and thermal expansion effects \cite{giustino2017electron,
malloy1991thermal}. Substantial thermal renormalisation of optoelectronic properties has been reported across a wide range of semiconductors,
from molecular crystals \cite{monserrat2015giant} to inorganic materials \cite{monserrat2018role,liu2023strong}, highlighting the fundamental
importance of finite-temperature effects in determining device-relevant properties.

An especially revealing class of materials in this context are strongly anharmonic perovskite-like systems, which can exhibit exceptionally 
large temperature-driven changes in their electronic structure. Among these, chalcohalide anti-perovskites (CAP) \cite{benitez2025giant}, 
exemplified by Ag$_3$SBr \cite{palazon1,palazon2}, display giant band-gap reductions of approximately $20$–$60$\% at room temperature 
relative to their zero-temperature values. Such behaviour arises from the pronounced anharmonicity of their lattice dynamics and the strong 
coupling between electronic and vibrational degrees of freedom. Beyond thermal stimuli, the combination of structural inversion symmetry and 
polar low-energy phonon modes in CAP has led to the proposal that external electric fields could be used to induce rapid and sizable 
optoelectronic changes \cite{benitez2025band}, offering an alternative route to dynamically control of band gaps and light-absorption 
features.

Recent experimental work has further expanded the scope of CAP by demonstrating chemically synthesised solid solutions of the form 
Ag$_{3}$SBr$_{1-x}$I$_{x}$ with highly promising optoelectronic properties \cite{cano2024novel}. These materials combine narrow band 
gaps around $1.0$~eV with strong absorption across the visible spectrum, while retaining considerable compositional flexibility through 
Br/I alloying. Their highly anharmonic lattice dynamics, temperature-dependent electronic structure, and coexistence of photoactivity 
with significant ionic mobility position CAP solid solutions as a particularly rich platform for exploring multifunctional behaviour in 
semiconductors. Together, these findings point to an unusually high degree of tunability in CAP optoelectronics and raise fundamental 
questions regarding the microscopic mechanisms governing their finite-temperature properties.

From a theoretical perspective, however, CAP present formidable challenges. Their strong anharmonicity renders conventional harmonic 
or quasi-harmonic descriptions inadequate, requiring an explicit treatment of electron–phonon coupling beyond these approximations, 
as well as thermal expansion effects \cite{benitez2025crystal}. While \textit{ab initio} methods such as density functional theory (DFT) 
and \textit{ab initio} molecular dynamics (AIMD) have successfully uncovered key aspects of CAP behaviour, they are computationally 
demanding and scale poorly with chemical disorder and reduced symmetry. As a result, the systematic modelling of CAP solid solutions, 
and, more broadly, of complex low-symmetry perovskite-inspired materials, remains largely out of reach, limiting both fundamental 
understanding and rational materials design.

\begin{figure*}
    \centering
    \includegraphics[width=1\linewidth]{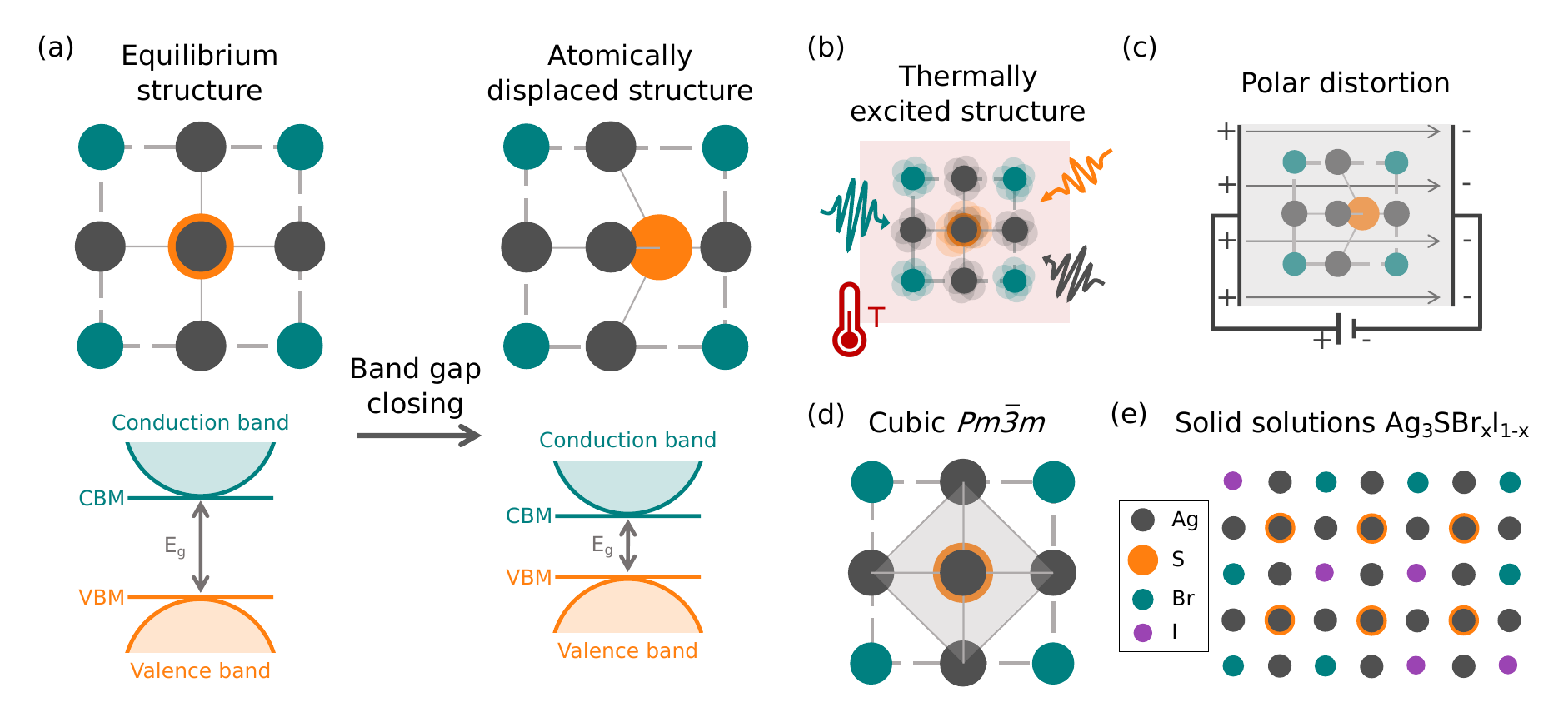}
    \caption{\textbf{Band-gap response to external stimuli in CAP}. 
    (a)~Band-gap closing as a result of structural perturbations condensing in the equilibrium cubic CAP structure. 
    Types of structural perturbations that may render band-gap closing in CAP, (b)~general thermal fluctuations, 
    and (c)~a polar distortion stabilised by an electric field. (d)~Cubic anti-perovskite structure of reference
    with space group $Pm\overline{3}m$. (e)~CAP solid solutions with chemical formula Ag$_3$SBr$_x$I$_{1-x}$.}
    \label{fig1}
\end{figure*}

In this study, we address this challenge by introducing a combined machine-learning (ML) and first-principles framework designed to 
efficiently capture anharmonic lattice dynamics and electron–phonon coupling in chemically disordered solids. We illustrate the 
capabilities of this approach on the technologically relevant Ag$_{3}$SBr$_{x}$I$_{1-x}$ system. By integrating machine-learning 
interatomic potentials (MLIP) \cite{jacobs2025practical,Batatia2022mace} with predictive graph neural networks (GNN) models 
\cite{paszke2019pytorch} trained on DFT data, we achieve near first-principles accuracy for energies, forces, stresses, and band 
gaps at a fraction of the computational cost. This approach enables a systematic and physically transparent exploration of thermal 
effects on vibrational stability and optoelectronic properties in low-symmetry materials, providing a new theoretical pathway for 
understanding and ultimately controlling dynamically reconfigurable semiconductor functionalities.

\section{Results and Discussion}
\label{sec:results}
We focus on Ag-based chalcohalide anti-perovskite (CAP) solid solutions with general chemical formula Ag$_3$SBr$_x$I$_{1-x}$ 
(Fig.~\ref{fig1}a). These materials can be synthesised via a variety of physical and chemical routes \cite{palazon1,palazon2,cano2024novel}, 
making them experimentally accessible and well suited for systematic investigation. Parent CAP are characterised by a giant 
band-gap reduction with increasing temperature (Fig.~\ref{fig1}b), a behaviour that originates from their pronounced lattice 
anharmonicity and strong electron–phonon coupling \cite{benitez2025giant,benitez2025band}. This combination of properties makes 
CAP-based compounds a particularly compelling testbed for the solid-solution modelling formalism introduced in this work.

In addition to thermal effects, CAP are predicted to exhibit band-gap variations of comparable magnitude under the application of 
external electric fields (Fig.~\ref{fig1}c). In this case, the underlying mechanism involves the condensation of polar phonon modes 
from the reference centrosymmetric cubic $Pm\overline{3}m$ phase (Fig.~\ref{fig1}d), highlighting the intimate coupling between lattice 
dynamics and electronic structure. Here, we investigate Ag$_3$SBr$_x$I$_{1-x}$ solid solutions (Fig.~\ref{fig1}e) by explicitly 
considering the stoichiometries $x~\in~\left\{0, 0.125, 0.25, 0.375, 0.5, 0.625, 0.75, 0.875, 1 \right\}$.

\begin{figure*}
    \centering
    \includegraphics[width=1\linewidth]{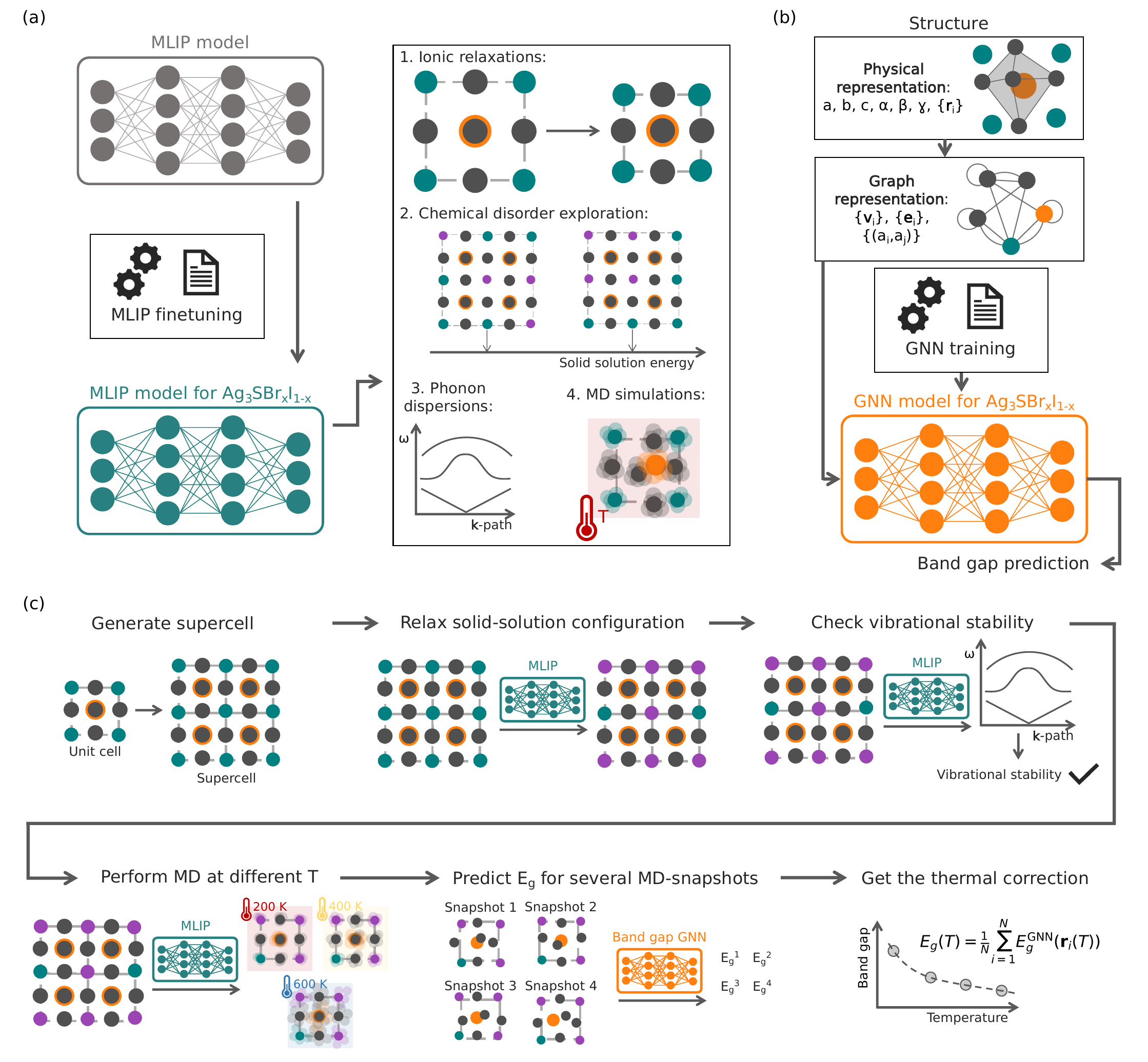}
    \caption{\textbf{Computational workflow for predicting temperature-dependent band gaps in solid solutions}.
    (a)~MLIP fine-tuning performed on a DFT-PBEsol dataset of energies, forces and stresses.
    (b)~Band-gap prediction workflow: the graph representation of a given atomic configuration is first obtained and subsequently
    processed by a GNN predictive model trained on a DFT-HSEsol dataset.
    (c)~General workflow for accurate band-gap prediction in solid solutions, integrating thermal corrections.}
    \label{fig2}
\end{figure*}

\subsection{General workflow}
\label{subsec:general}
We begin by constructing a comprehensive DFT dataset for Ag$_3$SBr$_{x}$I$_{1-x}$ solid solutions using both semilocal (PBEsol 
\cite{perdew2008restoring}) and range-separated hybrid (HSEsol \cite{schimka2011improved}) exchange–correlation functionals. 
Accurate prediction of temperature-dependent band gaps in strongly anharmonic materials requires explicit treatment of 
electron–phonon coupling \cite{benitez2025giant,benitez2025band}, which in practice entails sampling atomically displaced 
configurations representative of thermal fluctuations \cite{benitez2025physics}. Accordingly, the dataset comprises total 
energies, atomic forces, stresses, and electronic band gaps evaluated on such thermally perturbed structures (additional 
details are provided in the following section).

While PBEsol enables efficient evaluation of structural and vibrational properties, its well-known self-interaction errors 
limit the reliability of band-gap predictions \cite{garza2016predicting}, thereby necessitating hybrid-functional accuracy. 
However, for chemically disordered solid solutions with large and low-symmetry simulation cells, direct hybrid-DFT calculations 
become computationally prohibitive. To overcome this limitation, we adopt a dual machine-learning strategy (Fig.~\ref{fig2}). 
The PBEsol dataset is used to finetune a pre-trained universal MLIP \cite{Batatia2022mace}, which is subsequently employed 
for ionic relaxations, chemical-disorder exploration, phonon dispersion calculations, and finite-temperature molecular dynamics 
(MD) simulations (Fig.~\ref{fig2}a). In parallel, the HSEsol band-gap dataset is used to adjust a GNN model, initially trained 
on the PBEsol dataset, capable of reproducing hybrid-DFT band gaps with high fidelity (Fig.~\ref{fig2}b and Methods). This 
blended approach enables the explicit treatment of chemical disorder, anharmonic lattice dynamics, and hybrid-level electronic 
properties within a computationally tractable framework.

The combination of the fine-tuned MLIP and accurate GNN model enables a unified workflow for evaluating finite-temperature band-gap 
renormalisation in CAP solid solutions (Fig.~\ref{fig2}c). Starting from the cubic $Pm\overline{3}m$ antiperovskite unit cell, we 
construct supercells sufficiently large to accommodate each targeted stoichiometry $x$. Chemical disorder is then sampled using the 
MLIP. When the number of symmetry-inequivalent configurations for a given composition is manageable (considered here as $\lesssim 
100$ due to computational affordability reasons), all configurations are retained for further analysis. For compositions with 
a much larger configurational space ($> 1,000$), about a hundred of representative structures are generated using a random sampling 
procedure; all the resulting configurations are subsequently relaxed using the MLIP, and ranked according to their equilibrium energies.

The relaxed lowest-energy configurations are subsequently assessed for vibrational stability by computing its phonon frequencies 
with the MLIP. Only vibrationally stable structures are retained for the final stage of the analysis, in which temperature-dependent 
band gaps are evaluated. Long MD simulations are performed at selected temperatures using the MLIP, and electronic band gaps are 
computed for statistically uncorrelated MD snapshots using the GNN model. The ensemble-averaged band gaps obtained in this manner 
(Methods, Fig.~\ref{fig2}c) provide estimates of the temperature-renormalised band gaps of CAP solid solutions. In the following 
sections, we describe in detail the generation of the DFT datasets, the training of the GNN model, and the fine-tuning of the 
MLIP, together with a critical analysis of their performance. 

\begin{figure*}
    \centering
    \includegraphics[width=1\linewidth]{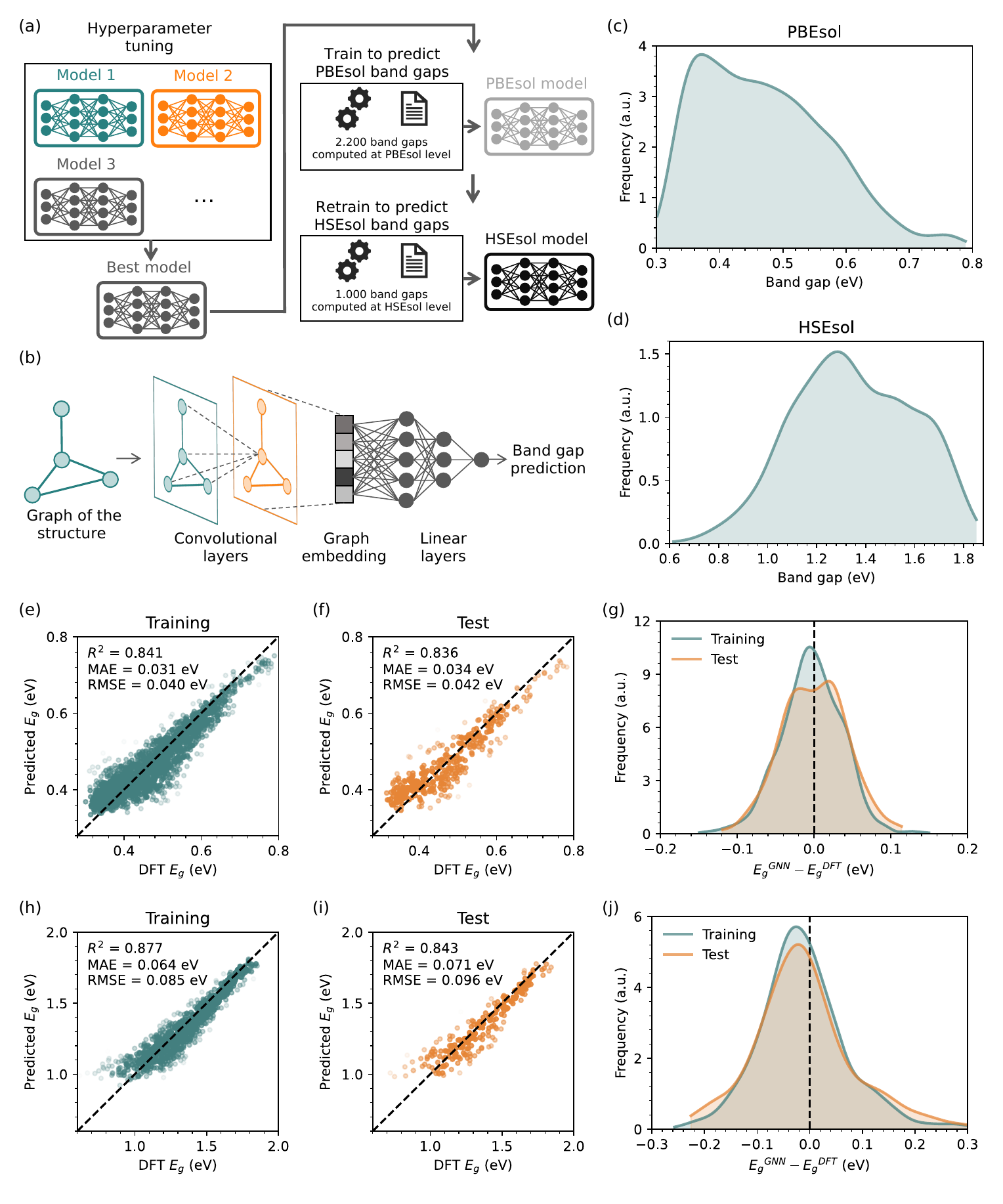}
    \caption{\textbf{GNN training for accurate band-gap prediction}.
    (a)~Workflow for training the GNN model on the DFT-HSEsol dataset.
    (b)~GNN model architecture.
    (c,d)~Band-gap distribution in the DFT-PBEsol and DFT-HSEsol training datasets.
    Predicted band gap vs. DFT calculated band gap for the training~(e) and test~(f) DFT-PBEsol datasets.
    Predicted band gap vs. DFT calculated band gap for the training~(h) and test~(i) DFT-HSEsol datasets.
    Band gap error distribution for GNN band-gap prediction on the DFT-PBEsol~(g) and DFT-HSEsol~(j) datasets.
    }
    \label{fig3}
\end{figure*}

\subsection{Dataset generation}
\label{subsec:dataset}
The datasets used for training the GNN band-gap model and for fine-tuning the MLIP were generated by introducing temperature-dependent 
phononic displacements, $\mathbf{u}^{\rm phon}$, into the reference antiperovskite cubic $Pm\overline{3}m$ structure 
\cite{benitez2025physics}. Phonon dispersions for the parent CAP compounds Ag$_3$SBr and Ag$_3$SI were computed using the small-displacement 
method as implemented in the \verb!Phonopy! package \cite{phonopy-phono3py-JPCM,phonopy-phono3py-JPSJ}. Atomic fluctuations about 
equilibrium positions were evaluated within the quasi-harmonic approximation \cite{qha1,qha2,qha3} according to the expression:
\begin{eqnarray}
        &       u_{j,\alpha}^{\rm phon}  =  \epsilon \cdot \zeta_{j,\alpha} \nonumber \\
        &  \zeta_{j,\alpha}^{2}  =  \frac{\hbar}{2Nm_{j}} \sum_{\mathbf{q},\nu} \frac{1}{\omega_{\nu}(\mathbf{q})}
        \left( 1 + 2n_{\nu}(\mathbf{q}, T) \right) \left| e_{\nu}^{\alpha} (j, \mathbf{q}) \right|^{2}
\label{eq:phon}
\end{eqnarray}
where $\hbar$ is the reduced Planck constant, $N$ is the total number of atoms in the unit cell, $m_{j}$ is the atomic mass of atom 
$j$, $\mathbf{q}$ denotes a reciprocal-space vector, $\alpha$ stands for Cartesian direction, $\nu$ is a phonon mode branch index, 
and $\omega_{\nu}$, $n_{\nu}$, and $e_{\nu}$ correspond to the phonon frequency, Bose-Einstein occupation distribution, and phonon 
eigenvector, respectively. The prefactor $\epsilon$ is a random number sampled from the uniform distribution $\mathcal{U}[-1,1]$, 
ensuring stochastic sampling of vibrational amplitudes.

All configurations were generated using a $2 \times 2 \times 2$ supercell containing $40$ atoms. Chemical disorder was introduced 
through atomic substitutions consistent with the general formula Ag$_3$SBr$_x$I$_{1-x}$, including the limiting compositions $x=0$ 
and $x=1$. Phononic displacements were applied according to Eq.~(\ref{eq:phon}) for temperatures between $100$ and $600$~K. In addition, 
a small number of equilibrium configurations were included to ensure that the dataset retained explicit information on the ideal lattice 
geometry; these latter configurations were used in the GNN training but not in the MLIP fine-tuning.

For each displaced configuration, total energies, atomic forces, mechanical stresses, and electronic band gaps were computed using 
the semilocal PBEsol exchange-correlation functional \cite{perdew2008restoring}, resulting in a total $2,200$ configurations. A further 
$1,000$ configurations were evaluated using the range-separated hybrid HSEsol functional \cite{schimka2011improved} to obtain high-accuracy 
band-gap data for re-training a GNN model (see next section).
 
\begin{figure*}
    \centering
    \includegraphics[width=1\linewidth]{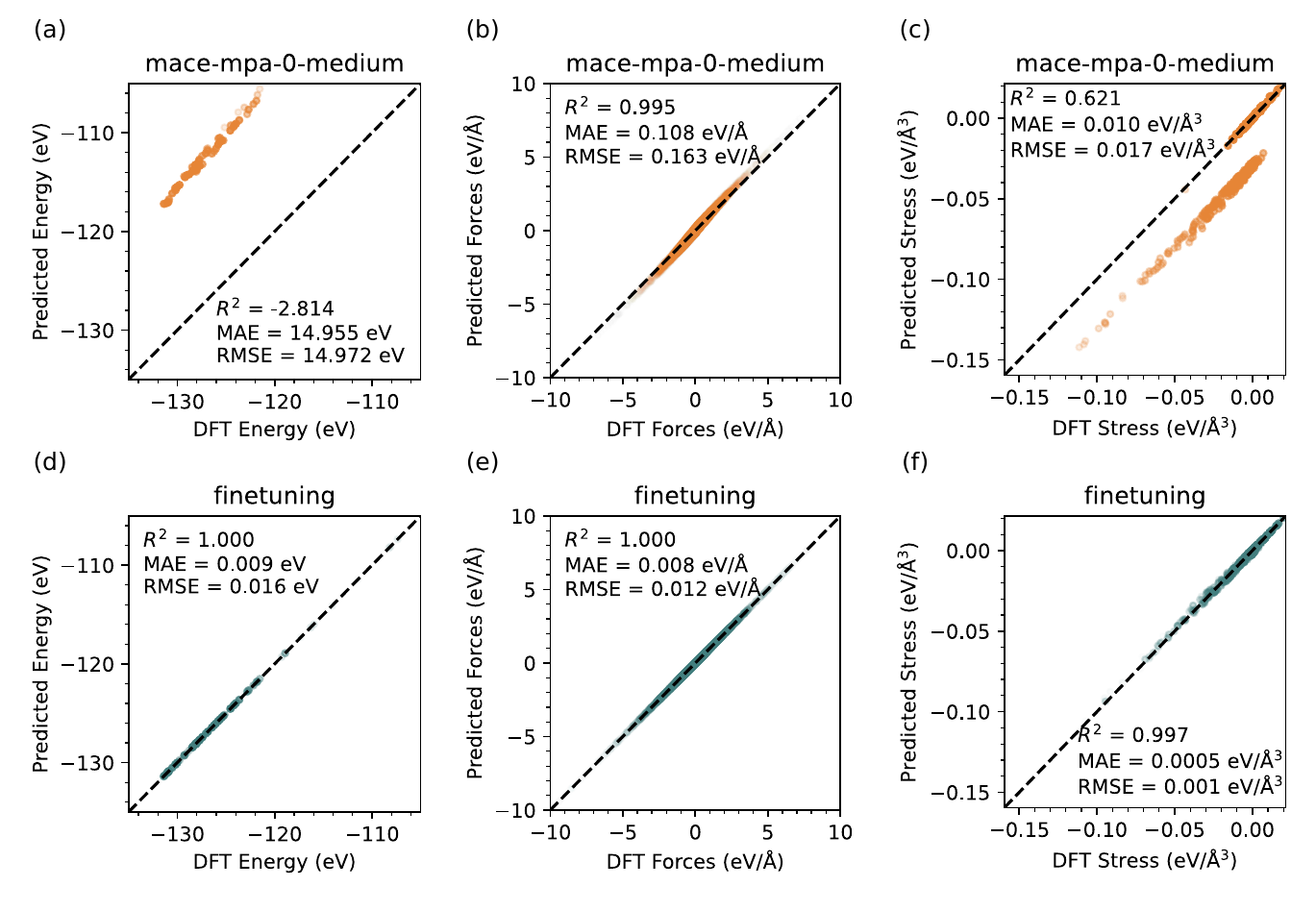}
    \caption{\textbf{MLIP fine-tuning for CAP solid solutions.}
            DFT vs. MLIP prediction of (a)~energies, (b)~atomic forces, and (c)~mechanical stress using the pre-trained MACE model 
	    \cite{Batatia2022mace}, and (d-f)~the fine-tuned MLIP.}
    \label{fig4}
\end{figure*}

\subsection{GNN model} 
\label{subsec:gnn}
To ensure robust predictive performance of the GNN band-gap model, we carried out a systematic exploration of the hyperparameter space 
(Fig.~\ref{fig3}a). This search involved testing multiple GNN architectures and training parameters under a common validation protocol, 
ultimately identifying the optimal model configuration (see Methods and work \cite{benitez2025physics} for technical details). As illustrated 
in Fig.~\ref{fig3}a, the progressive refinement of model architectures converges toward a single ``best model'', indicating that predictive 
performance is sensitive to architectural choices but stabilises once an adequate representational capacity is achieved.

The adopted architecture is schematically shown in Fig.~\ref{fig3}b. The model operates on graph representations of the atomic structures, 
where nodes encode atomic species and local descriptors, and edges represent bonding environments. Through successive graph convolutional 
(message-passing) layers, node features are iteratively updated using information from neighbouring atoms weighted by edge attributes. 
This mechanism enables the model to learn how local chemical environments influence electronic structure. A global pooling operation then 
generates a fixed-size graph embedding, ensuring that structures with different numbers of atoms and bonds can be processed consistently. 
The embedding is subsequently passed through fully connected layers to produce a scalar output corresponding to the electronic band gap.

Training was performed in two sequential stages. The model was first trained on the larger PBEsol dataset and subsequently fine-tuned on 
the smaller HSEsol dataset (Sec.~\ref{subsec:dataset}). The band-gap distributions for both datasets are shown in Figs.~\ref{fig3}c and d.
The PBEsol dataset exhibits a higher density of small band-gap values ($E_g \le 0.6$~eV), reflecting the systematic band-gap underestimation 
of semilocal functionals. In contrast, the HSEsol dataset is shifted toward larger gap values, as expected for hybrid functionals. The 
separation between these distributions underscores the importance of the second training stage: it enables the model to correct for the 
intrinsic bias of semilocal DFT while retaining the structural diversity learned from the larger dataset. Importantly, the two datasets 
contain distinct atomic configurations, so the HSEsol fine-tuning step introduces genuinely new structural information rather than reusing 
previously seen geometries.

The performance of the model after the first and second training stages is summarised in Figs.~\ref{fig3}e--g and h--j, respectively. The 
predicted versus DFT-computed band gaps for the training sets (Figs.~\ref{fig3}e and h) and test sets (Figs.~\ref{fig3}f and i) reveal a 
strong linear correlation, with only minor deviations at the extremes of the band-gap range. The error distributions (Figs.~\ref{fig3}g 
and j) are narrowly centred around zero, indicating the absence of significant systematic bias. Notably, the second training stage preserves 
high predictive accuracy despite the smaller dataset size, demonstrating the effectiveness of the transfer-learning strategy from PBEsol 
to HSEsol data. In both cases, the model achieves test-set $R^{2}$ values above $0.8$ and mean absolute errors (MAE) below $0.1$~eV. Given 
that the intrinsic accuracy of DFT band-gap predictions is typically of the order of $0.1$~eV, the GNN model therefore operates at near 
first-principles accuracy for Ag$_3$SBr$_{x}$I$_{1-x}$ solid solutions.

\begin{figure*}
    \centering
    \includegraphics[width=1\linewidth]{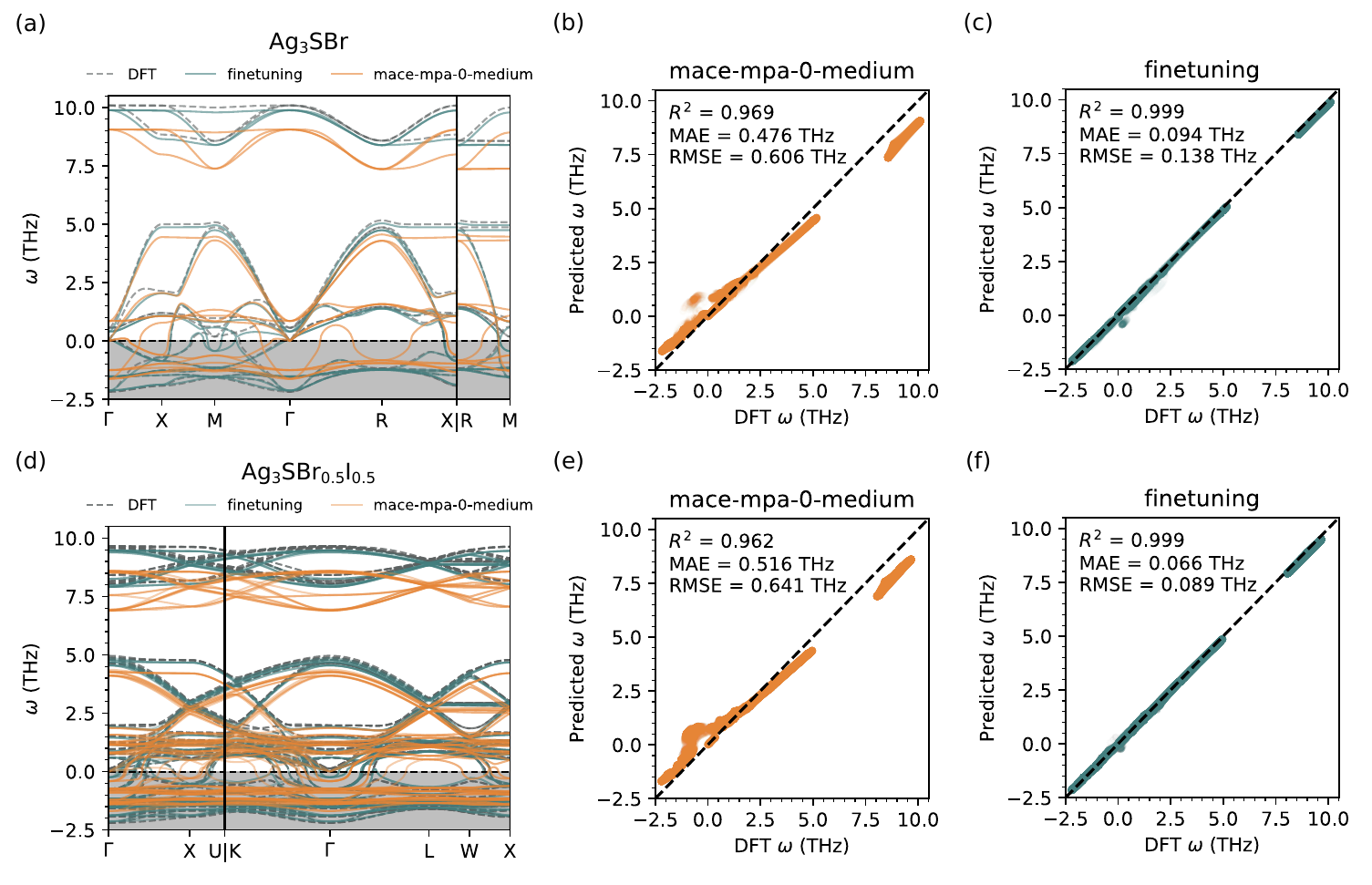}
    \caption{\textbf{Finetuned MLIP tested on phonon dispersions.} 
            (a)~Phonon dispersions calculated for Ag$_3$SBr using DFT, a pre-trained MACE model \cite{Batatia2022mace}, and the fine-tuned 
            MLIP. MLIP predictions vs. DFT calculations for (b)~a pre-trained MACE model, and (c)~the fine-tuned MLIP. 
            (d-f)~Equivalent phonon results obtained for a Ag$_3$SBr$_{0.5}$I$_{0.5}$ solid solution.}
    \label{fig5}
\end{figure*}

\subsection{MLIP fine-tuning} 
\label{subsec:mlip}
As a foundational machine-learning interatomic potential (MLIP), we selected the MACE framework in the \verb!mace-mpa-0-medium! 
flavour \cite{Batatia2022mace,Batatia2022Design,batatia2025foundation}. The MACE model was subsequently fine-tuned on our 
PBEsol dataset comprising total energies, atomic forces, and stress tensors in order to achieve a quantitatively accurate description 
of CAP solid solutions (Methods). The predictive performances of the original and fine-tuned models are compared in Fig.~\ref{fig4}.

Both models provide highly accurate atomic force predictions, with $R^2$ values close to unity (Figs.~\ref{fig4}b,e), indicating 
that the pre-trained foundational model captures the local bonding physics reasonably well. However, fine-tuning leads to a clear 
quantitative improvement, reducing the mean absolute error (MAE) in forces to $0.008$~eV/\AA. More substantial gains are 
observed in the prediction of total energies and stress tensor components. The re-trained MLIP achieves MAE values of $0.009$~eV 
for energies and $0.0005$~eV/\AA$^3$ for stresses (Figs.~\ref{fig4}d,f), demonstrating an accurate and internally consistent 
description of both structural and mechanical properties.

The apparently poor energy performance of the pre-trained model (Fig.~\ref{fig4}a) is largely attributable to a systematic offset 
between DFT and MACE reference energies, likely originating from differences in the exchange–correlation functional used during 
pre-training relative to PBEsol \cite{perdew2008restoring}. Once this offset is corrected through fine-tuning, the energy predictions 
align closely with the DFT values. In contrast, the pre-trained model shows limited transferability in predicting mechanical stress, 
as reflected by a relatively modest $R^2 = 0.621$ and MAE of $0.010$~eV/\AA$^3$ (Fig.~\ref{fig4}c). This highlights the importance 
of system-specific re-training when accurate elastic and thermomechanical properties are required.

To further validate the physical reliability of the fine-tuned MLIP, we assessed its ability to reproduce zero-temperature phonon 
dispersions for Ag$_{3}$SBr and Ag$_{3}$SBr$_{0.5}$I$_{0.5}$ (Fig.~\ref{fig5}). Although phonon frequencies were not explicitly 
included in the training dataset, accurate reproduction of DFT phonons is essential to justify the use of the MLIP for ionic 
relaxations and finite-temperature MD simulations. Figures~\ref{fig5}a,d compare harmonic phonon dispersions obtained from DFT 
with those predicted by the pre-trained and fine-tuned MLIP models, while Figs.~\ref{fig5}b,e and Figs.~\ref{fig5}c,f provide direct 
frequency-by-frequency comparisons.

The pre-trained \verb!mace-mpa-0-medium! model reproduces the overall qualitative features of the DFT phonon spectra, confirming 
that the foundational model captures the dominant interatomic interactions. Nevertheless, moderate frequency shifts are observed 
in both the low- and high-frequency regimes for Ag$_3$SBr and Ag$_3$SBr$_{0.5}$I$_{0.5}$ (Figs.~\ref{fig5}a,d), indicating incomplete 
transferability to CAP chemistry. By contrast, the fine-tuned model yields phonon dispersions that are nearly indistinguishable 
from the DFT results, including an accurate description of imaginary branches associated with lattice instabilities. The direct 
comparisons in Figs.~\ref{fig5}c,f show $R^2$ values close to unity and MAE values below $0.1$~THz, demonstrating quantitative 
agreement across the full vibrational spectrum.

Taken together, these results confirm that fine-tuning the MACE foundational model on a modest but physically diverse PBEsol dataset 
yields a highly accurate and transferable MLIP. The model reliably reproduces structural energetics, mechanical stresses, and lattice 
dynamics, thereby providing a robust foundation for the large-scale relaxations and finite-temperature simulations required to 
investigate CAP solid solutions.

\begin{figure*}
    \centering
    \includegraphics[width=1\linewidth]{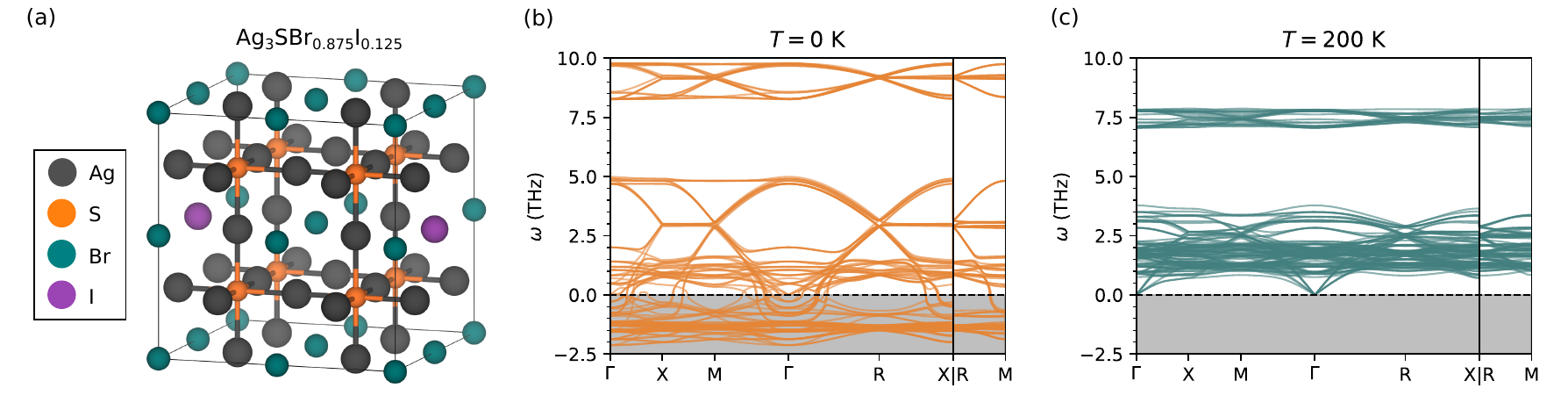}
    \caption{\textbf{Solid solution vibrational stability.} 
            (a)~Representation of the equilibrium structure determined for Ag$_3$SBr$_{0.875}$I$_{0.125}$ with the fine-tuned MLIP 
            model. Phonon dispersions calculated at (b)~$T = 0$~K, and (c)~at $T = 200$~K with the fine-tuned MLIP model.}
    \label{fig6}
\end{figure*}

\begin{figure*}
    \centering
    \includegraphics[width=1\linewidth]{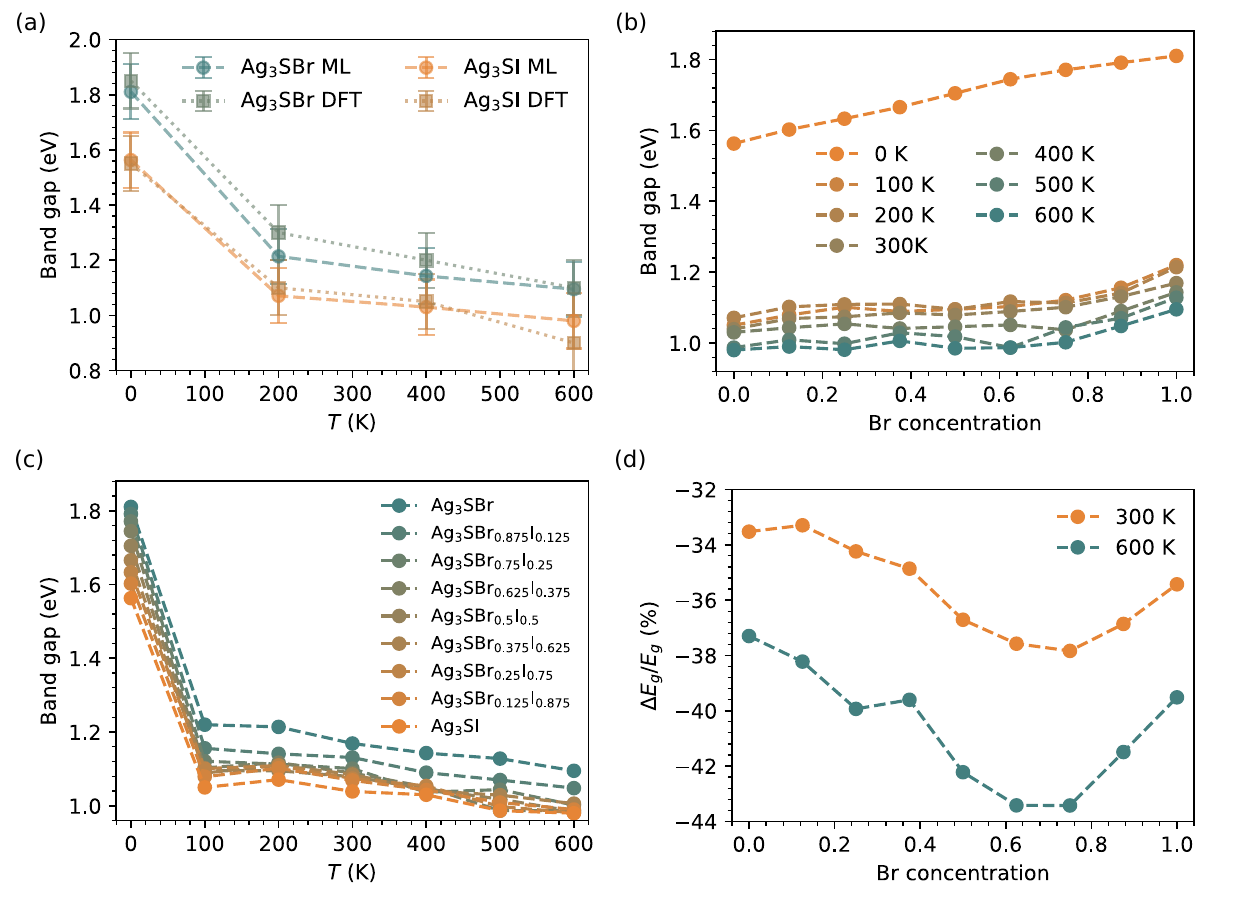}
    \caption{\textbf{Temperature-renormalised band gaps of CAP solid solutions.} 
            (a)~$T$-dependent band gaps obtained for Ag$_3$SBr and Ag$_3$SI with the DFT-HSEsol method and with the ML-based approach
	    introduced in this work. 
            (b)~Band gap of CAP solid solutions expressed as a function of Br concentration at different temperatures;
            error bars amount to $0.1$~eV, not shown to avoid overcrowding. 
            (c)~Band gap of CAP solid soltuions expressed as a function of temperature at different Br concentrations;
            error bars amount to $0.1$~eV, not shown to avoid overcrowding. 
            (d)~Relative variation of the band gap with respect to $T = 0$~K as a function of Br concentration at two 
                different temperatures.}
    \label{fig7}
\end{figure*}

\begin{figure*}
    \centering
    \includegraphics[width=1\linewidth]{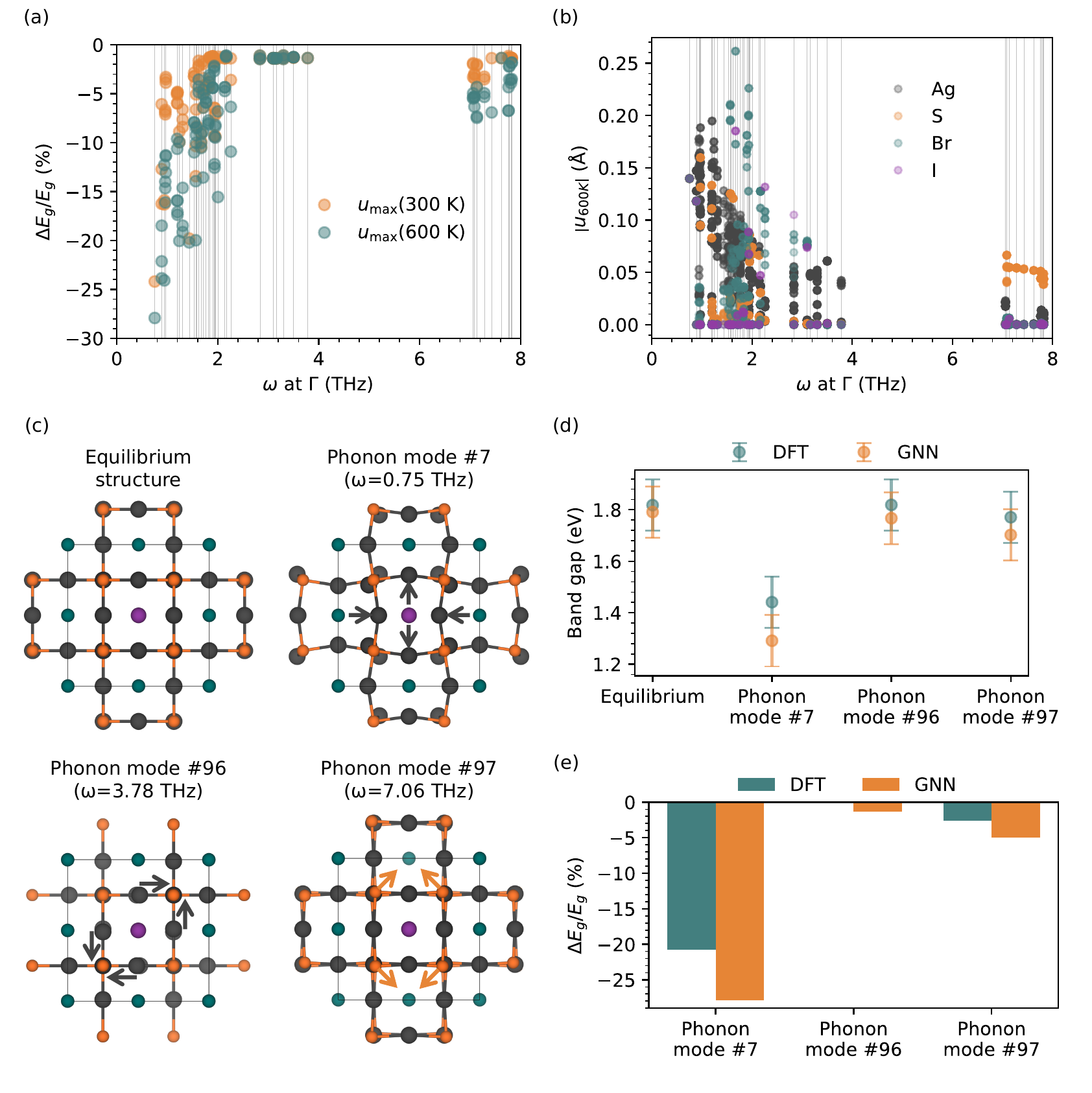}
    \caption{\textbf{Electron-phonon coupling mechanisms in Ag$_3$SBr$_{0.875}$I$_{0.125}$.} 
            (a)~Relative band-gap variation under $\Gamma$-phonon mode atomic displacements at $T = 300$ and $600$~K. 
            (b)~Atomic displacements corresponding to $\Gamma$-phonon modes calculated at $T = 600$~K. 
            (c)~Representation of the equilibrium structure and different $\Gamma$-phonon eigenmodes. Arrows indicate most relevant 
            atomic displacements. 
            (d)~Comparison between the DFT calculated and GNN predicted band gaps for the four atomic configurations represented in (c). 
            (e)~Relative band-gap change with respect to the equilibrium structure for specific $\Gamma$-phonon distortions.}
    \label{fig8}
\end{figure*}

\begin{table}[!htbp]
    \centering
    \begin{tabular}{ccccc}
    \hline
    \hline
            & & & &  \\
\quad Composition \quad & $E_{g}^{\rm exp}$~(300~K) &  \quad $E_{g}^{0K}$ & $E_{g}^{300K}$ & $E_{g}^{600K}$  \\
    & & & & \\
    \hline
    & & & & \\
     x = 0.0 & 0.9 & 1.6 & 1.0 & 1.0 \\
     x = 0.3 & 1.0 & 1.6 & 1.1 & 1.0 \\
     x = 0.5 & 1.0 & 1.7 & 1.1 & 1.0 \\
     x = 0.7 & 1.0 & 1.8 & 1.1 & 1.0 \\
     x = 1.0 & 1.0 & 1.8 & 1.1 & 1.0 \\
     & & & & \\
    \hline
    \hline
    \end{tabular}
\caption{{\bf Calculated and experimentally measured band gaps for CAP solid solutions.}
          Experimental values are extracted from previous work \cite{cano2024novel}. Both the experimental and theoretical 
          uncertainties approximately amount to $0.1$~eV. Results are expressed in units of eV.}
    \label{tab:experiments}
\end{table}

\subsection{Temperature-dependent optoelectronic properties of CAP solid solutions}
\label{subsec:bandgap}
Having established the accuracy of both the fine-tuned MLIP and predictive GNN band-gap model, we proceed to investigate the 
optoelectronic properties of chemically disordered Ag$_3$SBr$_x$I$_{1-x}$ solid solutions with $x \in \left\{0, 0.125, 0.25, 0.375, 0.5, 
0.625, 0.75, 0.875, 1 \right\}$. To maintain computational affordability while preserving configurational diversity, we employ a 
$40$-atom simulation cell constructed as a $2 \times 2 \times 2$ supercell of the cubic $Pm\overline{3}m$ unit cell. In this 
representation, Br/I substitutions occur over eight lattice sites, resulting in a manageable number of symmetry-inequivalent 
configurations. For example, at $x = 0.5$, a total of 70 distinct configurations arise, all of which can be explicitly simulated.

For each composition, all symmetry-inequivalent configurations were fully relaxed using the fine-tuned MLIP and ranked according to 
their equilibrium energies. The lowest-energy configuration at each stoichiometry was selected for subsequent lattice-dynamical 
analysis. Phonon dispersion relations were computed both at $T = 0$~K and under finite-temperature conditions. This step is 
essential because CAP systems are strongly anharmonic, and their vibrational stability can depend sensitively on temperature 
\cite{benitez2025crystal,benitez2025giant,benitez2025band}. Finite-temperature phonons were obtained using a normal-mode decomposition 
technique based on MD trajectories \cite{sun2014dynamic} (Methods), allowing anharmonic effects to be explicitly incorporated.

Figure~\ref{fig6} illustrates representative phonon dispersions for Ag$_3$SBr$_{0.875}$I$_{0.125}$, shown at $T = 0$~K (Fig.~\ref{fig6}b) 
and at $T = 200$~K (Fig.~\ref{fig6}c). At zero temperature, the presence of imaginary phonon branches indicates vibrational instability. 
Upon increasing temperature, however, these unstable modes are dynamically stabilised, and the spectrum becomes entirely real. This 
thermally induced stabilisation is consistent with previous findings for the parent compounds Ag$_3$SBr and Ag$_3$SI 
\cite{benitez2025crystal,benitez2025giant}. Supplementary Figs.~1–6 demonstrate that analogous behaviour occurs across the full 
compositional range considered here, confirming that dynamic stabilisation is a general feature of Ag$_3$SBr$_x$I$_{1-x}$ solid solutions. 
These results further highlight the central role of anharmonicity in governing the structural stability of CAP materials.

After establishing vibrational stability at finite temperature, we evaluated the temperature renormalisation of the electronic band gap. 
As a first validation of our ML-based framework, we applied it to the parent compounds Ag$_3$SBr and Ag$_3$SI, for which accurate 
first-principles and experimental data are available \cite{benitez2025giant}. The calculated band gaps at 0, 200, 400, and 600~K show 
excellent agreement, within the numerical uncertainties, with previous DFT-AIMD results \cite{benitez2025giant} (Fig.~\ref{fig7}a). Since 
those first-principles results are themselves in quantitative agreement with experiment \cite{benitez2025giant} (Table~1), this comparison 
confirms that the present ML-driven approach retains first-principles accuracy while enabling significantly enhanced computational 
efficiency.

The temperature-renormalised band gaps for all investigated compositions are summarised in Figs.~\ref{fig7}b,c as functions of both 
Br concentration and temperature. At zero temperature, the band gap increases systematically with increasing Br content (Fig.~\ref{fig7}b), 
reflecting the intrinsic compositional tuning of the electronic structure. This overall trend persists at finite temperatures, although 
minor non-monotonic variations appear in the intermediate range $0.4 \le x \le 0.6$, suggesting a subtle interplay between chemical 
disorder and lattice dynamics.

For any fixed composition, the band gap decreases with increasing temperature, with the largest reductions occurring in the low-temperature 
regime (approximately $0 \le T \le 100$~K), where electron–phonon coupling effects are most pronounced (Figs.~\ref{fig7}b,c). The relative 
band-gap changes with respect to the zero-temperature values are shown in Fig.~\ref{fig7}d. Thermal renormalisation effects are maximal 
in the intermediate compositional range $x \approx 0.6$--0.8, reaching approximately $-38$\% and $-44$\% at 300 and 600~K, respectively. 
In contrast, the smallest relative reductions are observed for the parent compound Ag$_3$SI (approximately $-33$\% and $-37$\% at 300 
and 600~K), although even these values remain remarkably large in absolute terms.

Table~1 presents a comparison between our finite-temperature band-gap predictions for CAP solid solutions and the available experimental 
data reported in work \cite{cano2024novel}. The estimated numerical uncertainties are on the order of $0.1$~eV for both the experimental 
measurements and the calculations. When thermal effects are explicitly accounted for through the ML-based framework introduced in this work, 
the agreement between first-principles predictions and experiment is excellent across all investigated compositions, remaining fully within 
the corresponding error bars.

Overall, these results validate the accuracy, reliability, and robustness of our ML-enabled predictive framework for optoelectronic 
properties of solid solutions under realistic finite-temperature conditions. Furthermore, they demonstrate that chemical composition 
and temperature act as powerful and intrinsically coupled tuning parameters for controlling the optoelectronic response of CAP solid 
solutions. In this context, anharmonic lattice dynamics emerge as a key physical mechanism governing the interplay between structural 
disorder and electronic structure, thereby enabling fine evaluation of band-gap behavior.

\subsection{Electron-phonon coupling mechanisms}
\label{subsec:elecphon}
In addition to evaluating finite-temperature band-gap renormalisation, we used our ML-based framework to analyse in greater detail 
the microscopic origin of electron–phonon coupling in CAP solid solutions. Specifically, we quantified the band-gap variations induced 
by frozen $\Gamma$-point phonon distortions of the reference antiperovskite structure, following a strategy analogous to that previously 
applied to the parent compounds Ag$_3$SBr and Ag$_3$SI \cite{benitez2025band,benitez2025giant}.

Figure~\ref{fig8}a presents the relative band-gap change, referenced to the equilibrium value, computed at $T = 300$ and $600$~K for 
Ag$_3$SBr$_{0.875}$I$_{0.125}$ under distortions corresponding to different $\Gamma$-phonon modes. The results are plotted as a function 
of the phonon frequency $\omega$, while the associated atomic displacement patterns (evaluated at the highest temperature) are shown in 
Fig.~\ref{fig8}b. A clear frequency-dependent trend emerges: the magnitude of the band-gap renormalisation roughly decreases with increasing 
phonon frequency, indicating that low-frequency modes contribute most strongly to the optoelectronic response.

The largest relative band-gap reductions, reaching approximately $-30$\%, are concentrated in the low-frequency region ($\omega \le 
2$~THz), where the vibrational modes are dominated by Ag displacements (Fig.~\ref{fig8}b). In contrast, modes in the intermediate 
frequency range ($2.5$--$4.0$~THz), characterised by coupled Ag and Br/I motions, produce negligible band-gap variations. At higher 
frequencies ($\omega \ge 7$~THz), where S displacements dominate, moderate band-gap reductions of about $-10$\% are observed. This 
hierarchy of contributions closely mirrors the behaviour previously reported for Ag$_3$SBr and Ag$_3$SI \cite{benitez2025band}, 
reinforcing the conclusion that soft, Ag-dominated lattice vibrations play a central role in governing electron–phonon coupling 
in CAP materials.

To further validate the predictive accuracy of the GNN model, we performed explicit DFT calculations of band-gap variations induced 
by representative $\Gamma$-phonon distortions spanning the three frequency regimes identified above. The corresponding displacement 
patterns are illustrated in Fig.~\ref{fig8}c, with arrows highlighting the dominant atomic contributions. As shown in Figs.~\ref{fig8}d,e, 
the GNN predictions are in very good agreement with the DFT results, within numerical uncertainties. 

Supplementary Figs.~7--12 present analogous $\Delta E_{g}/E_{g}$ calculations for the remaining compositions investigated in this work. 
Across all compositions, the same qualitative picture emerges: low-frequency modes induce the largest band-gap renormalisation, intermediate 
modes have diminished impact, and high-frequency modes produce moderate changes. These results provide a direct and quantitative link 
between lattice dynamics and optoelectronic response in CAP solid solutions. More broadly, the approach demonstrates how machine-learning 
models trained on first-principles data can be used not only to reproduce finite-temperature properties but also to dissect mode-resolved 
electron–phonon coupling in strongly anharmonic materials.

\section{Conclusions}
\label{sec:conclusions}
In this work, we have introduced and validated a machine-learning-assisted first-principles framework for the quantitative 
modelling of electron–phonon coupling and anharmonic lattice dynamics in chemically disordered solid solutions. By combining a 
fine-tuned MACE-based machine learning interatomic potential with a graph neural network trained to reproduce hybrid-DFT band 
gaps, we achieved near first-principles accuracy for energies, forces, stresses, phonon dispersions, and temperature-renormalised 
electronic band gaps in Ag$_3$SBr$_x$I$_{1-x}$ antiperovskites. The approach enables explicit sampling of chemical disorder, 
finite-temperature vibrational effects, and mode-resolved electron–phonon coupling at a computational cost that would be 
prohibitive within a fully hybrid-DFT framework.

Applied to Ag$_3$SBr$_x$I$_{1-x}$ solid solutions, our method reveals (i)~systematic compositional tuning of the zero-temperature 
band gap, (ii)~strong and composition-dependent thermal renormalisation reaching up to $\sim 40$--45\% at elevated temperatures, 
and (iii)~a dominant contribution of low-frequency, Ag-dominated vibrational modes to the electron–phonon coupling mechanism. 
We further demonstrate that these systems undergo temperature-induced dynamical stabilisation, confirming the decisive role of 
anharmonic lattice effects in governing both structural and optoelectronic properties. The excellent agreement obtained with 
previous first-principles and experimental data for the parent compounds validates the quantitative robustness of the ML-driven 
strategy.

Beyond the specific case of CAP materials, the present formalism opens new perspectives for computational materials science. The 
methodology is directly transferable to other highly anharmonic and chemically complex systems, including hybrid halide perovskites, 
oxide and chalcogenide perovskites, thermoelectric materials, phase-change materials, and low-symmetry solid solutions used in 
photovoltaics and photocatalysis. In particular, it provides a practical route to investigate temperature-dependent band structures, 
vibrational stabilisation mechanisms, and mode-resolved electron–phonon interactions in systems where large supercells, chemical 
disorder, and hybrid-functional accuracy are simultaneously required.

More broadly, this work illustrates how machine learning can move beyond property interpolation and become a tool for physically 
interpretable modelling of coupled electronic and lattice degrees of freedom. By enabling systematic exploration of dynamically 
reconfigurable semiconductors, the framework paves the way for predictive design of materials whose optoelectronic response can 
be tuned through composition, temperature, strain, or external fields. We anticipate that such approaches will play an increasingly 
central role in addressing open problems related to anharmonicity-driven phase stability, thermal band-gap engineering, and 
multifunctional energy materials.

\section*{Methods}
\subsection*{DFT calculations}
DFT calculations \cite{cazorla2017simulation,blochl1994projector} were performed with the \verb!VASP! software 
\cite{kresse1993ab,kresse1996efficiency,kresse1996efficient} and semilocal PBEsol exchange-correlation functional \cite{perdew2008restoring} 
to calculate energies, atomic forces, mechanical stresses and band gaps. Wave functions were represented in a plane-wave basis set 
truncated at $550$~eV. For reciprocal-space Brillouin zone sampling, we selected dense k-point grids equivalent to that of $12\times 12 
\times 12$ for the $5$-atoms cubic antiperovskite unit cell. By using these parameters, we obtained zero-temperature energies converged 
to within $0.5$~meV per formula unit. For geometry relaxations, a force tolerance of $0.005$~eV~\AA$^{-1}$ was imposed in all the atoms. 
Electronic density of states and accurate band gaps were computed with the range-separated hybrid HSEsol functional 
\cite{schimka2011improved}.

\subsection*{Graph generation}
Graphs representing materials structures were generated by employing a radius cutoff method \cite{fung2021benchmarking}. Each atom in 
the unit cell is represented as a node with four different chemical features: atomic number, atomic mass, atomic radius, and 
electronegativity. A cutoff radius $R_{\text{cutoff}}$ defines the maximum distance between pairs of atoms to be considered. Pairs of 
atoms closer than $R_{\text{cutoff}}$, taken equal to $5.5$~\AA~ in this study, are connected by an edge with their Euclidean distance 
as the edge feature. Atoms not connected by edges can still \textit{communicate} via the message passing implemented in the GNN 
architecture. The features of nodes and edges are all of them normalized to take values between $0$ and $1$. The structure periodicity 
is ensured by constructing a supercell large enough to contain all cutoff spheres centered on each unit cell atom. Connections between 
unit cell atoms and their periodic pictures in the supercell are represented in the graph as a self-edge. Further details in the graph 
construction workflow are provided in work \cite{benitez2025physics}.

\subsection*{GNN technical parameters}
Graph Neural Networks (GNN) for band gap prediction were implemented using the PyTorch Geometric framework \cite{fey2019fast}, which is 
built on top of the PyTorch machine learning library \cite{paszke2019pytorch}. Different GNN architectures with different parameters were 
trained and we studied their performance on determining band gaps by doing a hyperparameter exploration. Further details in the 
hyperparameter study of different GNN architectures are provided in \cite{benitez2025physics}. 

The architecture of the best GNN band-gap predicting model consists of a first convolutional layer that receives the four features of 
each node and generates $64$ node features by using the edge feature, followed by a batch normalization and applying a ReLU activation 
function. A second convolutional layer is used receiving now $64$ node features, followed again by a batch normalization and ReLU. Then 
a global max pooling is applied followed by a dropout of $40\%$ of the features. The $64$ features are then passed to a linear layer 
that outputs $16$ features after applying a ReLU activation function. Finally, the 16 features are passed to a final linear layer that 
outputs a final feature after applying a sigmoid activation function. Band gaps are also normalized to be consistent with this last 
activation function.

\subsection*{MLIP fine-tuning details and calculations}
As a foundational MLIP framework, we chose the higher order equivariant message passing model MACE 
\cite{Batatia2022mace,Batatia2022Design,batatia2025foundation}. Our antiperovskites dataset generated with the PBEsol functional 
was used to finetune MACE, with 95$\%$ of the structures grouped in the training set and 5$\%$ in the validation set. The parameters 
used for the re-training amount to a total of $256$ equivariant messages, a distance cutoff of $8$~\AA, and an energy, atomic forces, 
and mechanical stress weights of $1$, $150$, and $10$, respectively. To assess the performance of our fine-tuned model, we compare our 
results with those obtained with the pre-trained universal MACE model \verb!mace-mpa-0-medium! \cite{Batatia2022mace}.

The fine-tuned MLIP model was used to perform single point calculations (energy and atomic forces), ionic relaxations, and MD 
simulations. All these calculations were conducted within the Atomic Simulation Environment \cite{ase-paper}. Ionic relaxations were 
performed by using the BFGS method with a convergence criterion of maximum force of $0.01$~eV/\AA. MD simulations were implemented by 
using a Langevin algorithm with a coefficient friction of $10^{-3}$~fs$^{-1}$, a time step of $1$~fs, and a total simulation time of 
$50$~ps. The velocities were initialized by a Maxwell-Boltzmann Distribution.

\subsection*{Anharmonic phonon calculations}
The \verb!DynaPhopy! software \cite{carreras2017dynaphopy} was used to calculate the anharmonic lattice dynamics (i.e., $T$-renormalized 
phonons) of the studied systems. A normal-mode-decomposition technique \cite{sun2014dynamic} was employed in which the atomic velocities 
$\mathbf{v}_{jl}(t)$ ($j$ and $l$ represent particle and Cartesian direction indexes) generated during fixed-temperature MD simulation 
runs were expressed like:
\begin{equation}
\textbf{v}_{jl} (t) = \frac{1}{\sqrt{N m_{j}}} \sum_{\textbf{q}s}\textbf{e}_{j}(\textbf{q},s) 
	e^{i \textbf{q} \textbf{R}_{jl}^{0}} v_{\textbf{q}s}(t)~,
\label{eq2}
\end{equation}
where $N$ is the number of particles, $m_{j}$ the mass of particle $j$, $\mathbf{e}_{j}(\mathbf{q},s)$ a phonon mode eigenvector 
($\mathbf{q}$ and $s$ stand for the wave vector and phonon branch), $\mathbf{R}_{jl}^{0}$ the equilibrium position of particle $j$, 
and $v_{\mathbf{q}s}$ the velocity of the corresponding phonon quasiparticle. 

The Fourier transform of the autocorrelation function of $v_{\mathbf{q}s}$ was then calculated, yielding the power spectrum:
\begin{equation}
G_{\textbf{q}s} (\omega) = 2 \int_{-\infty}^{\infty} \langle v_{\textbf{q}s}^{*}(0) v_{\textbf{q}s}(t) \rangle e^{i \omega t} dt~. 
\label{eq3}
\end{equation}
Finally, this power spectrum was approximated by a Lorentzian function of the form:
\begin{equation}
G_{\textbf{q}s} (\omega) \approx \frac{\langle |v_{\textbf{q}s}|^{2} \rangle}{\frac{1}{2} \gamma_{\textbf{q}s} 
        \pi \left[ 1 + \left( \frac{\omega - \omega_{\textbf{q}s}}{\frac{1}{2}\gamma_{\textbf{q}s}} \right)^{2}  \right]}~, 
\label{eq4}
\end{equation}
from which a $T$-renormalized quasiparticle phonon frequency, $\omega_{\textbf{q}s} (T)$, was determined as the peak position, and the 
corresponding phonon linewidth, $\gamma_{\mathbf{q}s} (T)$, as the full width at half maximum. These calculations were performed with 
the fine-tuned MACE model.

\subsection*{Thermal band-gap corrections}
The electron-phonon correction to the band gap was computed as the difference between the band gap at zero temperature for the equilibrium 
structures and the average band gap obtained from the MD simulation at a given $T$ \cite{benitez2025giant,benitez2025band}, that we can 
express with:
\begin{equation}
    \Delta E_{g}(T)=\lim_{t_{0} \to \infty} \int_{0}^{t_{0}}E_{g}^{\mathbf{R}(t)}dt - E_{g}(0)~,
\end{equation}
where $\mathbf{R}$ represents the positions of the atoms in the supercell at a given time $t$ of the molecular dynamics simulation. This 
expression can be numerically approximated as:
\begin{equation}
    \Delta E_{g}(T)=\frac{1}{N}\sum_{k=1}^{N}E_{g}\left( \left\{ \mathbf{R}_{k} (T) \right\} \right) - E_{g}(0)~,
\end{equation}
where the band gap is averaged over a finite number, $N$, of configurations. 

Determination of the band gap was performed by using the GNN trained model on the HSEsol dataset for Ag$_3$SBr$_{x}$I$_{1-x}$ solid 
solutions. The MD simulations from which the atomic configurations were extracted were performed with the fine-tuned-MACE model.

\section*{Supporting Information}
Structural representation and phonon dispersions of all solid solutions investigated in thus study; 
effect of $\Gamma-$phonon displacements on the band gap for all solid solutions; 
estimation of coercive electric fields for inducing phonon distortions.
\\

\section*{Data Availability} 
All relevant scripts, including model training workflows and data preprocessing routines, are freely accessible in the GitHub repository:
\url{https://github.com/polbeni/ML-thermal-optoelectronics}.
\\

\section*{Acknowledgments}
P.B. acknowledges support from the predoctoral program AGAUR-FI ajuts (2024 FI-1 00070) Joan Oró, which is backed by the 
Secretariat of Universities and Research of the Department of Research and Universities of the Generalitat of Catalonia, 
as well as the European Social Plus Fund. C.C. acknowledges support by MICIN/AEI/10.13039/501100011033 under the grants 
PID2023-146623NB-I00 and PID2023-147469NB-C21 and by the Generalitat de Catalunya under the grants 2021SGR-00343, 2021SGR-01519 
and 2021SGR-01411. Computational support was provided by the Red Española de Supercomputación under the grants FI-2025-1-0015,
FI-2025-2-0006, FI-2025-2-0028 and FI-2025-3-0004. This work is part of the Maria de Maeztu Units of Excellence Programme 
CEX2023-001300-M funded by MCIN/AEI (10.13039/501100011033).  
\\

\bibliographystyle{unsrt}

\end{document}